\documentclass{aastex63}
\usepackage{multirow}
\usepackage{subfigure}
\usepackage{threeparttable}
\usepackage{booktabs}
\usepackage{dcolumn}
\usepackage{longtable}
\usepackage{threeparttablex}
\usepackage{amsmath}

\shortauthors{Li et al.}

\begin{document}

\title{Magnetization Factors of Gamma-Ray Burst Jets Revealed by a Systematic Analysis of the Fermi Sample}

\correspondingauthor{He Gao}
\email{gaohe@bnu.edu.cn}

\author[0000-0002-0823-4317]{An Li}
\affiliation{Institute for Frontier in Astronomy and Astrophysics, Beijing Normal University, Beijing 102206, China}
\affiliation{Department of Astronomy, Beijing Normal University, Beijing 100875, China}

\author[0000-0002-3100-6558]{He Gao}
\affiliation{Institute for Frontier in Astronomy and Astrophysics, Beijing Normal University, Beijing 102206, China}
\affiliation{Department of Astronomy, Beijing Normal University, Beijing 100875, China}

\author{Lin Lan}
\affiliation{CAS Key Laboratory of Space Astronomy and Technology, National Astronomical Observatories, Chinese Academy of Sciences, \\ Beĳing 100101, China}

\author[0000-0002-9725-2524]{Bing Zhang}
\affiliation{Nevada Center for Astrophysics, University of Nevada Las Vegas, Las Vegas, NV 89154, USA}
\affiliation{Department of Physics and Astronomy, University of Nevada Las Vegas, Las Vegas, NV 89154, USA}

\begin{abstract}

The composition of gamma-ray burst (GRB) jets remained a mystery until recently. In practice, we usually characterize the magnetization of the GRB jets ($\sigma_0$) through the ratio between the Poynting flux and matter (baryonic) flux. With the increasing value of $\sigma_0$, magnetic energy gradually takes on a dominant role in the acceleration and energy dissipation of the jet, causing the proportion of thermal component in the prompt-emission spectrum of GRBs to gradually decrease or even be completely suppressed. In this work, we conducted an extensive analysis of the time-resolved spectrum for all \textit{Fermi} GRBs with known redshift, and we diagnose $\sigma_0$ for each time bin by contrasting the thermal and nonthermal radiation components. Our results suggest that most GRB jets should contain a significant magnetic energy component, likely with magnetization factors $\sigma_{0}\geq 10$. The value of $\sigma_{0}$ seems vary significantly within the same GRB. Future studies with more samples, especially those with lower-energy spectral information coverage, will further verify our results.

\end{abstract}


\section{Introduction} \label{sec:intro}

After extensive research over the past 50 yr, substantial advancements have been achieved in comprehending gamma-ray bursts (GRBs;
\cite{zhang2018}). Nonetheless, certain lingering questions persist, such as the exact proportion of magnetic energy within GRB jets.

In the very early picture, the magnetic field issue within GRB jets was not addressed, but rather it was assumed GRB outflows originate from an initially hot “fireball” composed of photons, electron/positron pairs and a small amount of baryons \citep{Paczynski1986, Goodman1986, 1990ApJ...365L..55S}. After an initially rapid acceleration phase under fireball thermal pressure, a significant fraction of the initial fireball thermal energy is converted to the kinetic energy of the jet \citep{1993ApJ...415..181M, 1993MNRAS.263..861P}. As the jet expands to the radius of the photosphere, the hot photons escape while the baryonic matter continues to expand to even greater distances, reconverting the kinetic energy into energetic particles through internal collisional dissipation (internal shock model; \citealt{1994ApJ...430L..93R}). Synchrotron (and possibly also synchrotron self-Compton) radiation by these particles gives rise to the observed nonthermal gamma-ray emission \citep{1994ApJ...432..181M, 1996ApJ...466..768T}. In this scenario, the GRB spectrum is expected to consist of a combination of thermal and nonthermal components. 

Equipped with both the Gamma-ray Burst Monitor (GBM; \citealt{2009ApJ...702..791M}) and the Large Area Telescope \citep{2009ApJ...697.1071A}, Fermi spacecraft broadened the observational range to cover 6-7 orders of magnitude in energy, thereby having a considerable advantage for a comprehensive study of spectral components within the prompt emission spectra of GRBs. Based on the Fermi sample, one speculation of the synthesized prompt emission spectrum for GRBs has been established, which may include three elemental spectral components \citep{2011ApJ...730..141Z}: (I) a nonthermal band component; (II) a quasithermal component; and (III) another nonthermal component extending to high energies, and the significance of different spectral components may vary among different GRBs. These results have also been confirmed by other satellite data, such as Compton Gamma Ray Observatory/BATSE data \citep{2016ApJ...819...79G}.

After more than a decade of observations, Fermi has detected a population of GRBs that exhibit pure nonthermal spectra across several orders of magnitude in energy \citep{2009Sci...323.1688A,2011ApJ...730..141Z}. The absence of photospheric emission poses a significant challenge to the “fireball” model \citep{zhang2011}. 
One solution to solve this problem is that GRB outflows should contain certain proportion of magnetic energy so that the contribution of the thermal component can be significantly reduced \citep{2002MNRAS.336.1271D,Zhang2009,zhang2011,2019ApJ...882..184L}. In this case, the total luminosity of GRB jets can be decomposed into the sum of the luminosity from hot components ($L_{h}$) and the luminosity from cold (Poynting flux) components ($L_{c}$), where the magnetization parameter at the central engine $\sigma_0$ can be defined as
\begin{equation}
 \sigma_0 \equiv \frac{L_{c,0}}{L_{h,0}}.
\end{equation}
In a hybrid jet scenario \citep{Gao2015}, if $\sigma_0\ll1$, one gets back to a hot fireball. With the increasing value of $\sigma_0$, magnetic energy gradually takes on a dominant role in the acceleration and energy dissipation of the jet, causing the nonthermal radiation component associated with magnetic dissipation to become stronger compared to the thermal radiation component. For moderate values of $\sigma_0$, thermal radiation becomes subdominant. For extremely large values of $\sigma_0$, the thermal component is completely suppressed, and only featureless nonthermal radiation can be observed. 

From an observational perspective, we do have sources whose spectrum are dominated by thermal components \citep{2009ApJ...706L.138A}, which should correspond to $\sigma_0\ll1$; we also have cases where the thermal component is subdominant, \citep{2011ApJ...727L..33G,2013ApJ...770...32G,2015ApJ...807..148G,2015ApJ...814...10G,2016ApJ...819...79G,2016ApJ...831L...8G,2017ApJ...846..138G,2012ApJ...757L..31A,2014ApJ...784...17B,2014Sci...343...51P}, which should correspond to moderate values of $\sigma_0$. The most prevalent cases are sources with featureless nonthermal radiation, which are expected to have extremely large values of $\sigma_0$ \citep{2009Sci...323.1688A,2011ApJ...730..141Z}.

From a theoretical perspective, \cite{Gao2015} developed an analytical method to quantify the spectrum properties for a hybrid GRB jet with both a hot fireball component and a cold Poynting flux component. They developed (1) a "bottom-up" approach to predict the temperature and luminosity of the photosphere emission for an arbitrary value of $\sigma_0$; (2) a “top-down” approach to diagnose $\sigma_0$ using the observational quantities of the energy spectrum, especially by contrasting the thermal and nonthermal radiation components. Many studies have used this method to successfully explain specific GRBs that show a superposed thermal component in the spectra,\footnote{There are also works applying similar methods to \cite{Gao2015}, which theoretically explained individual sources with thermal and nonthermal superposed spectra and constrained their magnetization factors, e.g. GRB 100724B \citep{2011ApJ...727L..33G} and GRB 130323A \citep{2013ApJ...770...32G}.} such as GRB 110721A \citep{Iyyani2016}, GRB 130310A \citep{Qin2021}, GRB 130518A \citep{Siddique2022}, GRB 211211A \citep{Chang2023}.

In this paper, we intend to systematically apply this method (the "top-down" approach) to a large sample of GRBs observed by \textit{Fermi}, estimating the exact value of $\sigma_{0}$ for bursts with a superposed thermal component or the lower limit of $\sigma_{0}$ for bursts with a featureless nonthermal component. Eventually, we plan to obtain the distribution of $\sigma_{0}$ for all the samples, enabling us to conduct a comprehensive investigation and gain insights into the various constituents comprising the GRB jet.

\section{Data Reduction and Sample Selection} \label{sec:dintro}

In this work, we use the data observed by the \textit{Fermi} GBM, which is downloaded from the Fermi Science Support Center's FTP site\footnote{https://heasarc.gsfc.nasa.gov/FTP/fermi/} and processed with GBM data tools v1.1.1. 3359 GRBs were detected by Fermi/GBM up to 2022 August 1. We use time-tagged event (TTE) data and response (RSP or RSP2) files for spectral analysis and simulation. The TTE data file records each photon's arrival time with 2 $\mu$s temporal resolution and which of the 128 energy channels the photon registered. The RSP file records the response matrix of the detector at the trigger time. The RSP2 file records a time sequence of response matrices around the trigger point; it is used when the spectrum slice is significantly distant from the trigger time, making the response matrix in the RSP file inapplicable.

The GBM consists of two sets of detectors: 12 sodium iodide (NaI) scintillators sensitive to the lower end of the energy range and two cylindrical bismuth germanate (BGO) scintillators sensitive to the higher energy range. For the data sample, we selected two of the strongest triggered TTE data files from the NaI detectors, covering the energy range from 8 to 900 keV, and the strongest triggered TTE data file from the BGO detectors, covering the energy range from 500 keV to 30 MeV. Each TTE data file is accompanied by a corresponding calibration file.

For the purpose of this study, an initial screening of the data sample was conducted based on two criteria: (1) availability of redshift information, as the calculation method of $\sigma_{0}$ requires the redshift of the source; and (2) the signal-to-noise ratio (SNR) within the $T_{90}$ range higher than 30. This criterion ensures sufficient photon counts for credible results. The SNR calculation equation is as follows:

\begin{equation}
{\rm SNR}=\frac{N_{\rm all}-N_{\rm bak}}{\sqrt{N_{\rm bak}}},
\end{equation}

where $N_{\rm all}$ represents the photon counts within the source time range, including both source and background photon counts. The background photon counts, $N_{\rm bak}$, are estimated using background trends outside the $T_{90}$ time range via polynomial interpolation. Ultimately, 87 GRBs were selected for further analysis.

Given the evolving nature of central engine properties, the value of $\sigma_0$ might vary at different stages of GRB prompt emission. Therefore, utilizing time-resolved spectra with the smallest possible time bin is crucial for accurately constraining $\sigma_0$. For the selected sample, each source was binned based on a criterion of an SNR higher than 30, consistent with the selection criteria above. Additionally, we recognized that relying solely on SNR for defining time intervals may lead to mixing peaks and valleys of the light curves at different energies. To address this, during the selection of time intervals for each source, a manual review was conducted, and if necessary, manual adjustments were made to avoid this issue. Consequently, a total of 87 GRBs were divided into 318 slices (see Table \ref{table_all} for details).

We then fit each slice with three different spectral models: nonthermal, hybrid (nonthermal + thermal), and thermal. For the nonthermal spectrum, we adopt either cutoff power-law (“CPL” in the table) function or Band function ("Band" in the table; \citealt{Band1993}), which can be expressed as 
\begin{equation}
 F(E)=Ae^{-(2+\alpha)E/E_{\rm peak}}(\frac{E}{100\ \rm keV})^{\alpha},
\end{equation}
and
\begin{equation}
 F(E)=\left\{
 \begin{array}{l}
 A(\frac{E}{100\ {\rm keV}})^{\alpha}e^{-(2+\alpha)E/E_{\rm peak}},\ {\rm if}\ E<\frac{(\alpha-\beta)E_{\rm peak}}{2+\alpha}, \\
 A(\frac{(\alpha-\beta)E_{\rm peak}}{(2+\alpha)100\ {\rm keV}})^{\alpha-\beta}e^{\beta-\alpha}(\frac{E}{100\ {\rm keV}})^{\beta},\ {\rm otherwise} \\
 \end{array}\right.
\end{equation}
respectively, where $\alpha$ and $\beta$ are low-energy and high-energy photon spectral indices, respectively. $E_{\rm peak}$ is the $\nu F \nu$ peak in keV.

For the hybrid spectrum, we combined nonthermal spectral function with a blackbody function ("BB" in the talbe) described as follows:
\begin{equation}
F(E)=A\frac{E^2}{e^{E/kT}-1}
\end{equation}
where $kT$ is the temperature in unit of keV. 

For the thermal spectrum, we adopt the multicolour blackbody model \citep{Peer2012}, which can be described as \citep{2018ApJ...866...13H} :
\begin{align}
    N(E) =& \frac{8.0525(m+1)K}{(\frac{T_{\rm max}}{T_{\rm min}})^{m+1}-1}(\frac{kT_{\rm min}}{\rm keV})^{-2}I(E), \\
    I(E) =& (\frac{E}{kT_{\rm min}})^{m-1}\int_{\frac{E}{kT_{\rm min}}}^{\frac{E}{kT_{\rm max}}}\frac{x^{2-m}}{e^x-1}dx \notag
\end{align}
where $kT$ is the temperature in unit of keV. The spectrum consists of a superposition of Planck functions in the temperature range from $T_{{\rm min}}$ to $T_{{\rm max}}$. $m$ is the power-law index of the Planck functions' distribution.

Here we use the Bayesian information criterion (BIC) to determine the best-fitting model. The BIC is a criterion used to evaluate the best-fitting model from a finite set of models, where the model with the lowest BIC value is considered preferred. The BIC can be defined as follows: ${\rm BIC} = -2\ln L + k\ln(n)$, where $k$ represents the number of model parameters, $n$ is the number of data points, and $L$ is the maximum value of the likelihood function for the estimated model.

Eventually, we have 225 slices (in 55 GRBs) that display a featureless nonthermal spectrum, 81 slices (in 30 GRBs) that display a hybrid spectrum and only 12 slices (in GRB 090902B and GRB 210610B) that display a multicolour blackbody spectrum. For all 318 slices, the best-fitting models are collected in Table \ref{table_all} with their parameters.

\section{$\sigma_{0}$ estimation} \label{sec:cal}

The fundamental properties of a GRB jet can be characterized by a series of parameters, including the initial radius ($r_0$) at which the jet emanates from the central engine, the total luminosity ($L_{w,0}$) of the jet, the dimensionless entropy ($\eta$) that defines average total energy (rest-mass energy plus thermal energy) per baryon in the hot component, and the magnetization parameter at the central engine ($\sigma_0$). After leaving the central engine, the jet will continuously expand and accelerate under the combined effects of thermal pressure and magnetic pressure until it reaches its maximum velocity $\Gamma_{c}\simeq \eta(1+\sigma_0)$ at the coasting radius $r_{c}$. After the jet exceeds a radius of $r_{\rm ph}$, the photon optical depth for Thomson scattering drops below unity so that photons previously trapped in the jet can escape. The observed temperature and flux for the photosphere emission are essentially determined by the temperature at $r_0$, which reads
\begin{equation}
T_0 \simeq \left(\frac{L_w}{4\pi r_0^2 a c (1+\sigma_0)}\right)^{1/4},
\end{equation} 
where $a=7.56 \times 10^{-15} {\rm erg~cm^{-3}~K^{-4}}$ is the radiation density constant and by the position of the photosphere, namely, whether the photospheric emission occurs during the acceleration phase or the coasting phase of the jet. Based on the relative magnitudes of $\eta$ and $\sigma_0$, as well as the different relationships between  $r_{\rm ph}$ and $r_{c}$, \cite{Gao2015} divided the parameter space of the central engine into different regimes:\footnote{The jet undergoes a rapid acceleration phase initially with $\Gamma \propto r$ until reaching the radius of $r_{\rm ra}$, which is defined by the larger one of the thermal coasting radius and the magnetosonic point. In \cite{Gao2015}, two extreme regimes with $r_{\rm ph} < r_{\rm ra}$ are also discussed. We do not consider these extreme scenarios here since they introduce an additional degeneracy so that central engine parameters cannot be inferred from observations.} (I) $\eta>(1+\sigma_{0})^{1/2}$ and $r_{\rm ph}<r_{c}$; (II) $\eta>(1+\sigma_{0})^{1/2}$ and $r_{\rm ph}>r_{c}$; (III) $\eta<(1+\sigma_{0})^{1/2}$ and $r_{\rm ph}<r_{c}$; and (IV) $\eta<(1+\sigma_{0})^{1/2}$ and $r_{\rm ph}>r_{c}$. For each regime, \cite{Gao2015} developed an analytical method to connect the observed quantities of GRB spectrum with the central engine parameters; for instance, they find that
\begin{equation}
    1+\sigma_{0}=\left\{
    \begin{array}{lr}
        \vspace{1ex}
        25.5(1+z)^{4/3}(\frac{kT_{\rm ob}}{50{\rm keV}})^{4/3}(\frac{F_{\rm BB}}{10^{-8}{\rm erg\ s^{-1}cm^{-2}}})^{-1/3}r_{0,9}^{2/3}f_{\rm th,-1}^{-1}f_{\gamma}^{-1}d_{\rm L,28}^{-2/3}, & \rm{Regime\ I} \\ \vspace{1ex}
        5.99(1+z)^{4/3}(\frac{kT_{\rm ob}}{30{\rm keV}})^{4/3}(\frac{F_{\rm BB}}{10^{-7}{\rm erg\ s^{-1}cm^{-2}}})^{-1/3}r_{0,9}^{2/3}f_{\rm th,-1}^{-1}f_{\gamma}^{-1}d_{\rm L,28}^{-2/3}, & \rm{Regime\ II \ and\ III} \\
        6.43(1+z)^{4/3}(\frac{kT_{\rm ob}}{10{\rm keV}})^{4/3}(\frac{F_{\rm BB}}{10^{-9}{\rm erg\ s^{-1}cm^{-2}}})^{-1/3}r_{0,9}^{2/3}f_{\rm th,-1}^{-1}f_{\gamma}^{-1}d_{\rm L,28}^{-2/3}, & \rm{Regime\ IV} \\
    \end{array}
\right.    
\end{equation}
where $T_{\rm ob}$ is the observed blackbody temperature and $F_{\rm BB}$ is the observed blackbody flux. $f_{\rm th}$ is defined as $F_{\rm BB}/F_{\rm ob}$, where $F_{\rm ob}$ is the observed total flux (both thermal and nonthermal included). $f_\gamma = L_\gamma/L_w$, which connects the total flux $F_{\rm ob}$ to the wind luminosity $L_w$. The $L_w$ is estimated by $L_w=4\pi d_L^2 F_{\rm ob}/f_{\gamma}$, in which $d_L$ is the luminosity distance of the GRB. 

\begin{figure}
    \centering
    \includegraphics{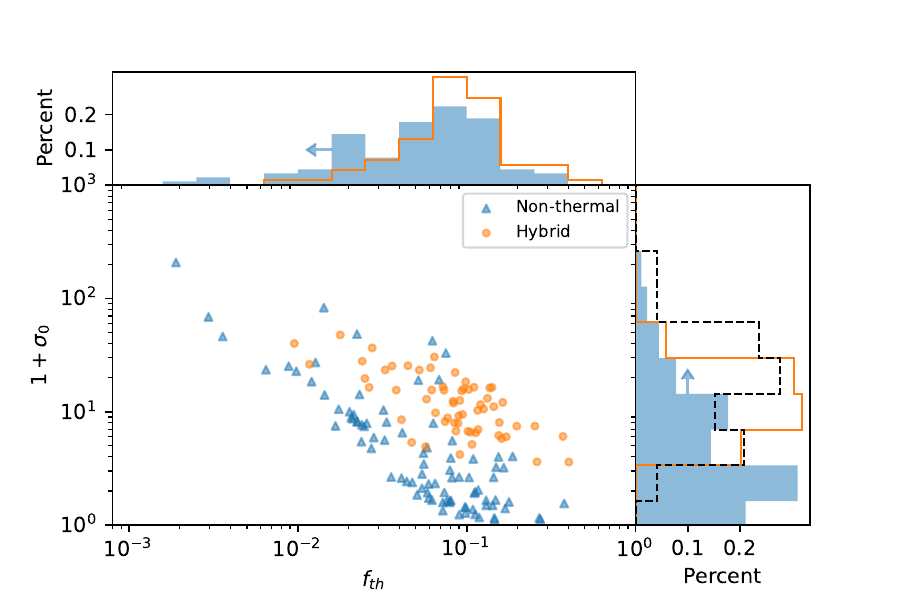}
    \caption{Distribution plot of $1+\sigma_{0}$ and $f_{\rm th}$. The blue triangles represent the lower limit of $1+\sigma_{0}$ for slices with non-thermal spectrum, the orange points represent the value of $1+\sigma_{0}$ for slices with hybrid spectrum. The black dashed line in the sub figure represents the distribution result of $1+\sigma_{0}$ taken from \cite{Li2019}. The results shown in this figure are based on the same $r_0$ value as $10^9$ cm.}
    \label{fig:sig_relation}
\end{figure}

Here we use the method proposed in \cite{Gao2015} to estimate the value of $\sigma_{0}$ for our selected sample. For slices that display a multicolour blackbody spectrum, we set their $\sigma_0$ equal to 0. For slices that display a hybrid spectrum, we first extract the observed quantities such as $T_{\rm ob}$, $F_{\rm BB}$ and $f_{\rm th}$ by fitting the spectrum. With these observed quantities, we can determine which regime these slices fall into by using the judgment criteria provided in Table \ref{table_eq1} (see also Table 1 and 2 in \cite{Gao2015}). Then, we can use Equation 8 to calculate the corresponding $\sigma_{0}$ for these slices. The results are shown in Figure \ref{fig:sig_relation} and Table \ref{table_all}. It is worth noting that in our estimation, we fix $f_\gamma$ to be a typical value of 0.5, and we consider three $r_0$ values: $10^8$cm, $10^9$cm, and $10^{10}$cm.

For slices with a featureless nonthermal component, we adopted the following approach to estimate their lower limit of $\sigma_{0}$:

\begin{itemize}
    \item For each slice, we first assume that its thermal component corresponds to a temperature of $kT$, with $kT$ starting at 5 keV.
    \item For a given $kT$, we assume a series of ratios of thermal component flux to nonthermal component flux ($f_{\rm th}$), with $f_{\rm th}$ values starting at 0.005 and increasing with equal steps (0.1 dex) on a log scale.
    \item For each pair of $kT$ and $f_{\rm th}$, we use the same response file as the original data to simulate the corresponding thermal component photons with random noise through the GBM data tools \citep{GbmDataTools}.
    \item We add these simulated photons to the original data, reperform spectral fitting, and compare the BIC results of the featureless nonthermal spectrum with the hybrid spectrum. If the nonthermal spectrum is the best fit, we continue to increase the $f_{\rm th}$ value and resimulate more thermal photons until the hybrid spectrum becomes dominant. This way, we have found the upper limit of $f_{\rm th}$ for a given $kT$.
    \item Taking $kT$ and the upper limit of $f_{\rm th}$ as the observed quantities, we use the judgment criteria provided in Table \ref{table_eq1} and equations 8 to calculate the corresponding $\sigma_{0}$, which could be taken as the lower limit of $\sigma_{0}$ corresponding to $kT$.
    \item We continuously increase the assumed value of $kT$ (from 5 keV to 60 keV with a uniform step size of 5 keV), calculate the corresponding lower limit of $\sigma_{0}$ for each $kT$ using the methods above, and take the minimum value of all the lower limits as the overall lower limit of $\sigma_{0}$ for this slice.
\end{itemize}

\begin{figure}
    \centering
    \includegraphics{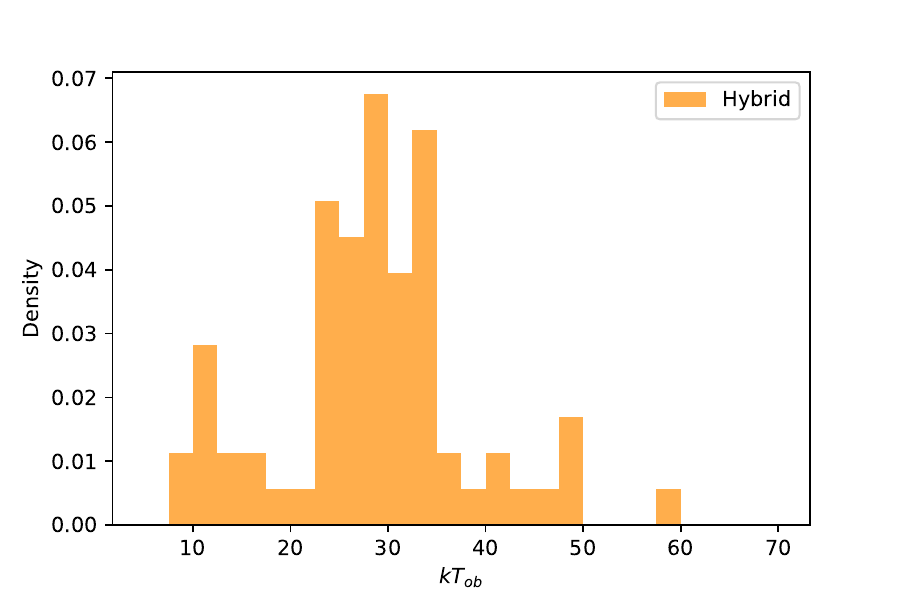}
    \caption{The distribution of $kT_{\rm ob}$ in the GRB with hybrid spectrum.}
    \label{fig:comp}
\end{figure}

\begin{figure}
    \centering
    \subfigure{\includegraphics[width=0.5\textwidth]{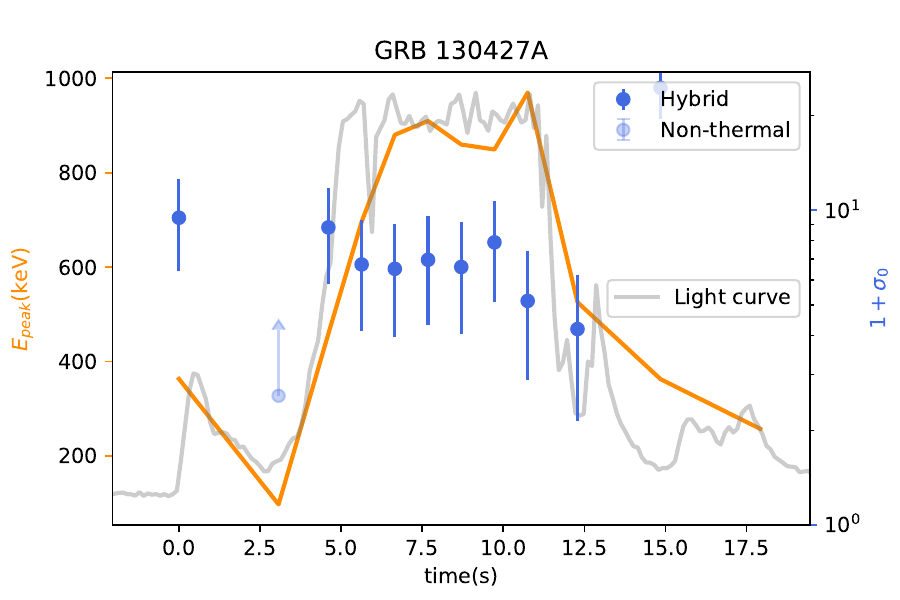}}
    \subfigure{\includegraphics[width=0.5\textwidth]{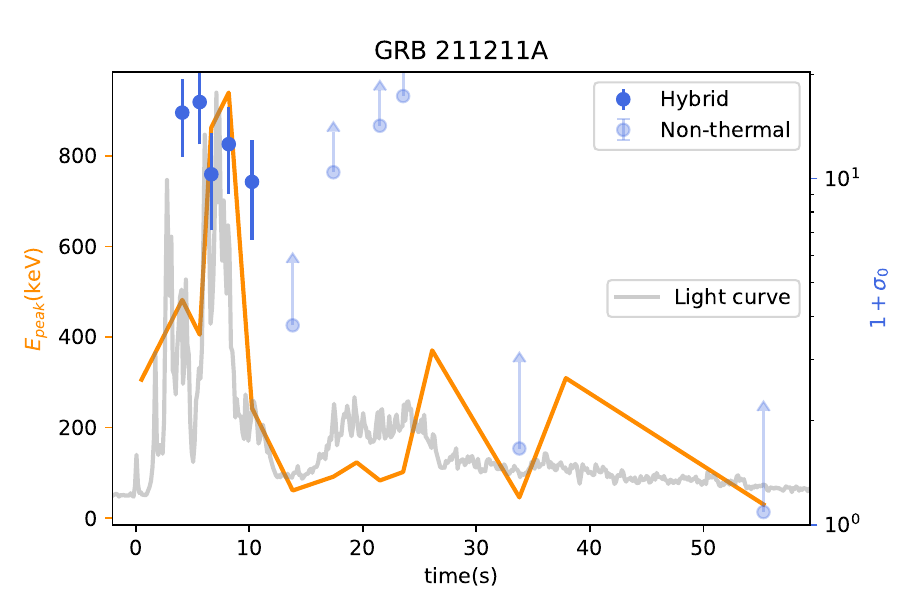}}
    \subfigure{\includegraphics[width=0.5\textwidth]{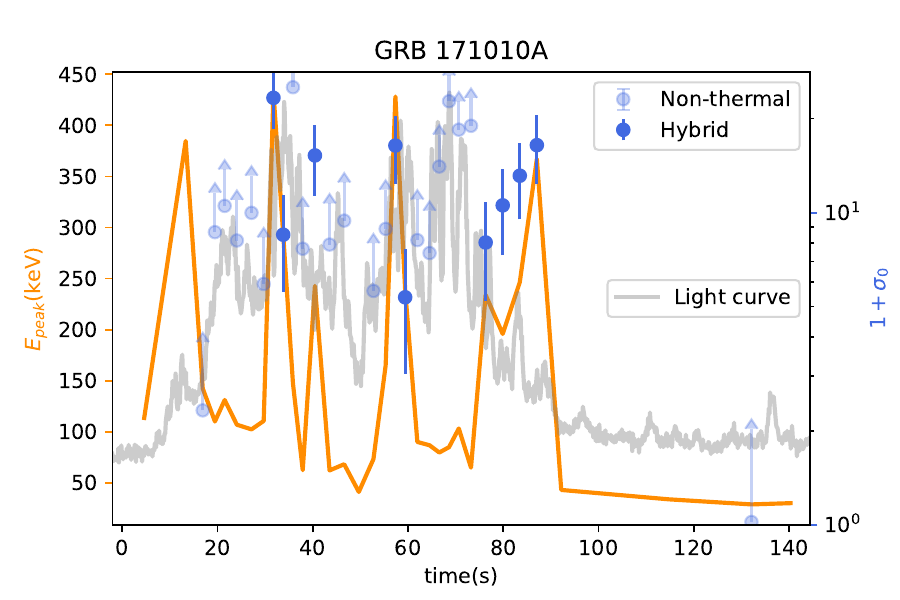}}
    \caption{Evolution of photon counts, $1+\sigma_{0}$ and $E_{\rm peak}$, for sources that show hybrid spectra in multiple slices.}
    \label{fig:example}
\end{figure}

In Figure \ref{fig:sig_relation}, we show the distribution plot of $1+\sigma_{0}$ vs. $f_{\rm th}$ for all slices with both nonthermal and hybrid spectra. For the hybrid slice, we find that $1+\sigma_{0}$ values are distributed in the range of 3.6-47.7, and if fitted with a Gaussian function, its peak is at 11.9 with a variance of 0.3. On the other hand, $f_{\rm th}$ is distributed in the range of 0.009-0.4, and if fitted with a Gaussian function on a log scale, its peak is at 0.08 with a variance of 0.3. Also, we show the distribution plot of $1+\sigma_{0}$ lower limit vs. $f_{\rm th}$ upper limit for all slices with a featureless nonthermal spectrum. We find that the $1+\sigma_{0}$ lower limits are distributed in the range of 1.1-424.9, with $\sim23\%$ larger than 10. The $f_{\rm th}$ upper limits are distributed in the range of 0.003-0.4. 

As shown in Figure \ref{fig:comp}, we find that the values of $kT_{\rm ob}$ for all slices with the hybrid spectrum are distributed in the range of 8.5-58.7 keV, with the average value located at 28.4 keV with a variance of 10.5 keV.

\section{conclusion and discussion} \label{sec:res}

In this study, we conducted an extensive analysis of the magnetic component within GRBs observed by the Fermi satellite. Our focus was on GRB data with known redshift, and we employed a time-resolved approach to determine the parameter $\sigma_{0}$ for each time bin. We categorized the GRBs into three distinct groups based on their dominant spectral components and applied different processing procedures for each:
\begin{itemize}
    \item Thermal-dominant spectra: in the case of 12 slices with thermal-dominant spectra, we set their $\sigma_0$ equal to 0.

    \item Hybrid spectra: we have 81 slices with hybrid spectra, for which we calculated their $\sigma_{0}$ values based on the extracted quantities such as $T_{\rm ob}$, $F_{\rm BB}$ and $f_{\rm th}$ by fitting the spectrum. 
    
    \item Nonthermal spectra: we have 225 slices with nonthermal spectra, for which we estimate the lower limit of their $\sigma_{0}$ value by simulating the upper limit of the thermal component.   
\end{itemize}

We find that for all slices with the hybrid spectrum, the values of $kT_{\rm ob}$ are distributed around 28 keV with a variance of 11 keV, and the values of $f_{\rm th}$ are distributed around 0.08 with a variance of 0.3 on a log scale. In this case, their $1+\sigma_{0}$ values should be distributed around 11.9 with a variance of 0.3 on a log scale. These results are consistent with previous results obtained by using a limited number of specific GRBs \citep{Zhang2009,Guiriec2011,Iyyani2013,Li2019,Chang2023}. On the other hand, we find that for all slices with the featureless nonthermal spectrum, the $1+\sigma_{0}$ lower limits are distributed in the range of 1.1-424.9, with $22.7\%$ larger than 10. 

Our results suggest that most GRB jets should contain a significant magnetic energy component, likely with magnetization factors $\sigma_{0}\geq 10$. In this case, the prompt emission mainly arises from magnetized dissipation in such highly magnetized environments as particle acceleration via internal shocks is suppressed. With detailed analysis of a small number of sources that show hybrid spectra in multiple slices, we find that the value of $\sigma_{0}$ varies significantly within the same GRB. In Figure \ref{fig:example}, we present three examples (GRB 171010A, GRB 211211A, and GRB 130427A) that best reflect the evolution of $\sigma_{0}$. These examples are selected because they have thermal components in three or more consecutive time bins. The evolution trend of $\sigma_{0}$ does not show an obvious relationship with those of lightcurve and $E_{peak}$.

We would like to point out that we have only considered the case where $kT$ varies from 5 to 60 keV when constraining the lower limit of the magnetization factor for the nonthermal spectra. The reasons are (1) the SNR of Fermi/GBM observations below 5 keV is too low to perform a good spectral fitting; and (2) according to the fitting results of the hybrid spectra slices, there are almost no cases where $kT$ is greater than 60 keV. For these nonthermal spectra, if their true $kT$ is greater than 60 keV, then the corresponding lower limit of the magnetization factor should be higher than our current results, which would further support our conclusion that GRB jets have high magnetization. However, if the true $kT$ is lower than 5 keV, then our current constraint on the lower limit of the magnetization factor may be overestimated to a certain extent. However, this uncertainty should be within a factor of 2. In the future, instruments like Fermi, Gravitational Wave High-energy Electromagnetic Counterpart All-sky Monitor, Einstein Probe \citep{2018SPIE10699E..25Y} and Space Variable Objects Monitor \citep{SVOM} are expected to form an observation network in a broadband and provide high-quality GRB data during the prompt emission and early-afterglow phases through joint analysis, which would significantly enhance the accuracy of our results.

Previous studies have indicated that if a more complex model is used to fit the energy spectrum, some slices that originally satisfy the pure nonthermal spectrum may show weaker evidence of the existence of a thermal spectrum (e.g., GRB 080916C and GRB 090926A; see \cite{2016ApJ...819...79G} for details). The fitting of the energy spectrum performed here does not include such complexity. However, it is important to emphasize that the lower limit of $\sigma_{0}$ given by the pure nonthermal spectrum in this study is the most conservative constraint. If thermal components are indeed missed, the constraint on $\sigma_{0}$ brought by these thermal components will definitely be above the lower limit we have provided, which would further strengthen our conclusion that GRB jets generally possess high magnetization.

Finally, we note that the low-energy end of the BGO detector was excluded from the analysis due to its low SNR. To ensure proper calibration between the NaI and BGO detectors, we also conducted analysis by including the low-energy end of the BGO data (e.g. over the energy range from 300 keV to 30 MeV), and we found that our main results were barely affected.

\acknowledgments

We gratefully acknowledge the insightful comments provided by the anonymous referee and Asaf Pe’er. This work is supported by the National Natural Science Foundation of China (Projects 12373040,12021003), the National SKA Program of China (2022SKA0130100), the National Postdoctoral Program for Innovative Talents (grant No. GZB20230765) and the Fundamental Research Funds for the Central Universities.

\vspace{5mm}
\facilities{Fermi(GBM)}

\software{astropy \cite{2013A&A...558A..33A},  
          GBM data tools \cite{GbmDataTools}
          }

\bibliography{export}{}
\bibliographystyle{aasjournal}

\begin{ThreePartTable}

\begin{TableNotes}
    \item[1] $\sigma_{0}$ is constrained in different $r_{0}$, the unit of $r_{0}$ is cm. As we mentioned in \ref{sec:dintro}, we choose $10^8$, $10^9$ and $10^{10}$ corresponding to the three different $r_{0}$ value in table: $r_{0,1}$, $r_{0,2}$ and $r_{0,3}$.
\end{TableNotes}

\begin{longtable}{c|c|c|c|c|c|c|c|c|c}

    \caption{$\sigma_{0}$ constraint result}
    \label{table_all} \\
    \hline
    Burst Name & Start Time$(T_0+s)$ & End Time$(T_0+s)$ & Fit Model & $E_{\rm peak}$ & $kT$ & $f_{\rm th}$ & \multicolumn{3}{c}{$1+\sigma_{0}$} \\
    \cline{8-10}
    & & & & (keV) & (keV) & & & & \\
    & & & & & & & $r_{0,1}$\tnote{1} & $r_{0,2}$\tnote{1} & $r_{0,3}$\tnote{1} \\
    \hline
    \endfirsthead

    \hline
    \hline
    Burst Name & Start Time$(T_0+s)$ & End Time$(T_0+s)$ & Fit Model & $E_{\rm peak}$ & $kT$ & $f_{\rm th}$ & \multicolumn{3}{c}{$1+\sigma_{0}$} \\
    \cline{8-10}
    & & & & (keV) & (keV) & & & & \\
    & & & & & & & $r_{0,1}$\tnote{1} & $r_{0,2}$\tnote{1} & $r_{0,3}$\tnote{1} \\
    \hline
    \endhead

    \hline
    \hline
    \endfoot
    
    \hline
    \hline
    \insertTableNotes
    \endlastfoot

    \hline
    \hline 
    \multirow{1}{*}{GRB080804A}
    & $-2.048$ & $15.360$ & CPL+BB & $209.948$ & $26.682$ & $0.1980$ & $1.609$ & $7.472$ & $ - $ \\ 
    \hline 
    \multirow{1}{*}{GRB080810A}
    & $-2.048$ & $22.528$ & CPL & $4265.090$ & $5.000$ & $0.3020$ & $ - $ & $ - $ & $2.314$ \\ 
    \hline 
    \multirow{1}{*}{GRB080916C}
    & $-2.048$ & $15.360$ & Band & $149.793$ & $5.000$ & $0.3546$ & $ - $ & $ - $ & $1.840$ \\ 
    \hline 
    \multirow{1}{*}{GRB081121A}
    & $-2.048$ & $16.384$ & Band & $129.227$ & $5.000$ & $0.1781$ & $ - $ & $ - $ & $3.719$ \\ 
    \hline 
    \multirow{1}{*}{GRB081221A}
    & $20.480$ & $24.576$ & CPL & $123.083$ & $20.000$ & $0.4317$ & $ - $ & $ - $ & $4.529$ \\ 
    \hline 
    \multirow{2}{*}{GRB081222A}
    & $-2.048$ & $6.144$ & Band & $147.967$ & $5.000$ & $0.1218$ & $ - $ & $1.189$ & $5.524$ \\ 
    & $6.144$ & $14.336$ & Band & $77.290$ & $5.000$ & $0.5013$ & $ - $ & $ - $ & $1.004$ \\ 
    \hline 
    \multirow{1}{*}{GRB090102A}
    & $10.240$ & $19.456$ & Band & $1130.951$ & $5.000$ & $0.1575$ & $ - $ & $ - $ & $3.349$ \\ 
    \hline 
    \multirow{3}{*}{GRB090323A}
    & $-2.048$ & $26.624$ & Band+BB & $261.177$ & $33.910$ & $0.1369$ & $3.500$ & $16.257$ & $ - $ \\ 
    & $26.624$ & $53.248$ & Band & $1060.073$ & $5.000$ & $0.1458$ & $ - $ & $1.155$ & $5.367$ \\ 
    & $53.248$ & $64.512$ & Band & $1021.761$ & $5.000$ & $0.1446$ & $ - $ & $ - $ & $4.049$ \\ 
    \hline 
    \multirow{3}{*}{GRB090328A}
    & $-2.048$ & $14.336$ & Band & $581.272$ & $5.000$ & $0.1686$ & $ - $ & $ - $ & $3.807$ \\ 
    & $14.336$ & $23.552$ & Band & $1287.037$ & $5.000$ & $0.1740$ & $ - $ & $ - $ & $3.038$ \\ 
    & $23.552$ & $38.912$ & Band & $545.817$ & $5.000$ & $0.3475$ & $ - $ & $ - $ & $1.521$ \\ 
    \hline 
    \multirow{2}{*}{GRB090424A}
    & $-2.048$ & $2.048$ & Band & $67.578$ & $5.000$ & $0.0595$ & $ - $ & $2.303$ & $10.698$ \\ 
    & $2.048$ & $5.120$ & Band+BB & $268.941$ & $9.271$ & $0.0563$ & $1.057$ & $4.906$ & $22.788$ \\ 
    \hline
    \multirow{1}{*}{GRB090510A}
    & $0.000$ & $1.000$ & CPL & $3864$ & $50.000$ & $0.0753$ & $7.107$ & $32.986$ & $ - $ \\ 
    \hline
    \multirow{1}{*}{GRB090516A}
    & $-7.168$ & $19.456$ & Band & $4287.681$ & $5.000$ & $0.3353$ & $ - $ & $ - $ & $2.420$ \\ 
    \hline
    \multirow{9}{*}{GRB090618A}
    & $48.128$ & $60.416$ & Band & $130.992$ & $10.000$ & $0.4676$ & $ - $ & $ - $ & $2.308$ \\ 
    & $60.416$ & $64.512$ & Band & $413.189$ & $5.000$ & $0.0611$ & $ - $ & $1.853$ & $8.607$ \\ 
    & $64.512$ & $67.584$ & Band & $156.887$ & $5.000$ & $0.0476$ & $ - $ & $2.379$ & $11.049$ \\ 
    & $67.584$ & $70.656$ & Band & $108.743$ & $5.000$ & $0.0587$ & $ - $ & $1.930$ & $8.964$ \\ 
    & $70.656$ & $74.752$ & Band & $70.328$ & $20.000$ & $0.1295$ & $1.109$ & $5.153$ & $ - $ \\ 
    & $74.752$ & $79.872$ & Band+BB & $286.238$ & $23.984$ & $0.1694$ & $1.297$ & $6.026$ & $ - $ \\ 
    & $79.872$ & $83.968$ & Band & $78.404$ & $5.000$ & $0.0640$ & $ - $ & $2.146$ & $9.970$ \\ 
    & $83.968$ & $88.064$ & Band & $98.230$ & $10.000$ & $0.1325$ & $ - $ & $1.932$ & $8.973$ \\ 
    & $88.064$ & $93.184$ & Band+BB & $48.621$ & $28.964$ & $0.1944$ & $ 1.337 $ & $ 6.212 $ & $ - $ \\ 
    \hline
    \multirow{7}{*}{GRB090902B}
     & $-2.048$ & $5.120$ & mBB & $5.000$ & $-0.012$ & $0.0481$ & $ - $ & $2.610$ & $12.125$ \\ 
     & $5.120$ & $8.192$ & mBB & $6.346$ & $-0.064$ & $0.0137$ & $2.856$ & $13.266$ & $61.577$ \\ 
     & $8.192$ & $10.240$ & mBB & $4.951$ & $-0.574$ & $0.0380$ & $ - $ & $2.101$ & $9.757$ \\ 
     & $10.240$ & $13.312$ & mBB & $20.000$ & $-1.147$ & $0.0482$ & $2.471$ & $11.480$ & $53.284$ \\ 
     & $13.312$ & $16.384$ & mBB & $10.000$ & $-0.752$ & $0.0423$ & $1.081$ & $5.019$ & $23.314$ \\ 
     & $16.384$ & $19.456$ & mBB & $5.000$ & $-0.775$ & $0.0509$ & $ - $ & $1.846$ & $8.575$ \\ 
     & $19.456$ & $22.528$ & mBB & $28.219$ & $-0.975$ & $0.0891$ & $1.992$ & $9.253$ & $ - $ \\ 
    \hline
    \multirow{4}{*}{GRB090926A}
    & $-2.048$ & $4.096$ & Band & $759.906$ & $5.000$ & $0.0572$ & $ - $ & $1.826$ & $8.480$ \\ 
    & $4.096$ & $7.168$ & Band & $198.750$ & $5.000$ & $0.0357$ & $ - $ & $2.657$ & $12.340$ \\ 
    & $7.168$ & $10.240$ & Band & $230.006$ & $10.000$ & $0.0834$ & $ - $ & $2.334$ & $10.843$ \\ 
    & $10.240$ & $13.312$ & Band & $187.977$ & $10.000$ & $0.1064$ & $ - $ & $1.829$ & $8.496$ \\ 
    \hline 
    \multirow{2}{*}{GRB091003A}
    & $-2.048$ & $10.240$ & Band & $680.186$ & $5.000$ & $0.1635$ & $ - $ & $ - $ & $3.758$ \\ 
    & $10.240$ & $20.480$ & Band & $288.675$ & $29.303$ & $0.0852$ & $3.819$ & $17.740$ & $ - $ \\ 
    \hline 
    \multirow{2}{*}{GRB091127A}
    & $-2.048$ & $2.048$ & Band & $46.689$ & $10.000$ & $0.0250$ & $4.944$ & $22.963$ & $106.585$ \\ 
    & $2.048$ & $8.192$ & Band & $19.917$ & $5.000$ & $0.0704$ & $ - $ & $3.272$ & $15.197$ \\ 
    \hline 
    \multirow{3}{*}{GRB100414A}
    & $-2.048$ & $10.240$ & CPL & $975.324$ & $5.000$ & $0.0821$ & $ - $ & $1.530$ & $7.106$ \\ 
    & $10.240$ & $18.432$ & CPL & $1730.237$ & $5.000$ & $0.0740$ & $ - $ & $1.541$ & $7.159$ \\ 
    & $18.432$ & $25.600$ & CPL & $2047.803$ & $5.000$ & $0.1245$ & $ - $ & $ - $ & $3.506$ \\ 
    \hline 
    \multirow{1}{*}{GRB100728A}
    & $-2.048$ & $22.528$ & CPL & $4287.280$ & $5.000$ & $0.2388$ & $ - $ & $ - $ & $2.680$ \\ 
    & $22.528$ & $48.128$ & CPL & $4288.484$ & $5.000$ & $0.2873$ & $ - $ & $ - $ & $2.228$ \\ 
    & $48.128$ & $61.440$ & CPL & $957.657$ & $10.000$ & $0.4029$ & $ - $ & $ - $ & $2.441$ \\ 
    & $61.440$ & $73.728$ & CPL & $1005.259$ & $10.000$ & $0.4366$ & $ - $ & $ - $ & $2.043$ \\ 
    & $73.728$ & $82.944$ & CPL & $230.883$ & $5.000$ & $0.1261$ & $ - $ & $ - $ & $4.186$ \\ 
    & $82.944$ & $91.136$ & CPL & $932.038$ & $5.000$ & $0.1321$ & $ - $ & $ - $ & $3.996$ \\ 
    & $91.136$ & $102.400$ & CPL & $411.993$ & $20.000$ & $0.4897$ & $ - $ & $1.023$ & $4.750$ \\ 
    & $102.400$ & $117.760$ & CPL & $173.773$ & $10.000$ & $0.4936$ & $ - $ & $ - $ & $2.195$ \\ 
    & $117.760$ & $132.096$ & CPL & $280.561$ & $10.000$ & $0.5062$ & $ - $ & $ - $ & $1.942$ \\ 
    & $132.096$ & $146.432$ & CPL & $185.942$ & $5.000$ & $0.2339$ & $ - $ & $ - $ & $3.016$ \\ 
    & $146.432$ & $171.008$ & CPL & $4288.458$ & $5.000$ & $0.3292$ & $ - $ & $ - $ & $1.944$ \\ 
    & $171.008$ & $186.368$ & CPL & $163.816$ & $10.000$ & $0.3202$ & $ - $ & $ - $ & $4.524$ \\ 
    \hline
    \multirow{1}{*}{GRB100814A}
    & $-2.048$ & $17.408$ & CPL+BB & $222.629$ & $30.438$ & $0.4026$ & $ - $ & $3.598$ & $ - $ \\ 
    \hline 
    \multirow{1}{*}{GRB100906A}
    & $-2.048$ & $12.288$ & Band & $74.296$ & $5.000$ & $0.1849$ & $ - $ & $ - $ & $2.855$ \\ 
    \hline 
    \multirow{1}{*}{GRB101213A}
    & $-2.048$ & $18.432$ & Band & $2485.779$ & $5.000$ & $0.2483$ & $ - $ & $ - $ & $3.121$ \\ 
    \hline 
    \multirow{1}{*}{GRB101219B}
    & $-2.048$ & $19.456$ & CPL & $4286.129$ & $5.000$ & $0.2915$ & $ - $ & $ - $ & $2.905$ \\ 
    \hline 
    \multirow{2}{*}{GRB110721A}
    & $-2.048$ & $3.072$ & Band+BB & $1334.871$ & $34.419$ & $0.0854$ & $4.784$ & $22.206$ & $ - $ \\ 
    & $3.072$ & $10.240$ & Band & $428.720$ & $50.000$ & $0.1595$ & $4.376$ & $ - $ & $ - $ \\ 
    \hline 
    \multirow{2}{*}{GRB110731A}
    & $-2.048$ & $5.120$ & Band+BB & $192.040$ & $47.753$ & $0.2533$ & $1.609$ & $7.473$ & $ - $ \\ 
    & $5.120$ & $9.216$ & Band & $1094.638$ & $5.000$ & $0.1450$ & $ - $ & $ - $ & $3.482$ \\ 
    \hline
    \multirow{2}{*}{GRB111228A}
    & $-12.288$ & $5.120$ & Band & $13.423$ & $5.000$ & $0.2635$ & $ - $ & $ - $ & $2.979$ \\ 
    & $12.288$ & $54.272$ & Band & $14.930$ & $5.000$ & $0.2000$ & $ - $ & $1.370$ & $6.365$ \\ 
    \hline 
    \multirow{3}{*}{GRB120119A}
    & $-2.048$ & $14.336$ & Band & $525.853$ & $10.000$ & $0.4796$ & $ - $ & $ - $ & $2.051$ \\ 
    & $14.336$ & $22.528$ & Band & $95.258$ & $5.000$ & $0.1446$ & $ - $ & $ - $ & $3.655$ \\ 
    & $22.528$ & $36.864$ & Band & $133.701$ & $10.000$ & $0.4273$ & $ - $ & $ - $ & $2.796$ \\ 
    \hline 
    \multirow{1}{*}{GRB120326A}
    & $-7.168$ & $9.216$ & Band & $36.329$ & $5.000$ & $0.5097$ & $ - $ & $ - $ & $1.037$ \\ 
    \hline 
    \multirow{1}{*}{GRB120624B}
    & $-2.048$ & $13.312$ & Band & $341.531$ & $5.000$ & $0.1411$ & $ - $ & $ - $ & $4.141$ \\ 
    \hline 
    \multirow{5}{*}{GRB120707A}
    & $-2.048$ & $12.288$ & CPL+BB & $245.757$ & $28.013$ & $0.1208$ & $2.492$ & $11.574$ & $ - $ \\ 
    & $12.288$ & $21.504$ & Band+BB & $52.935$ & $20.232$ & $0.1653$ & $ - $ & $ 4.368 $ & $ - $ \\ 
    & $21.504$ & $28.672$ & Band & $116.720$ & $5.000$ & $0.1336$ & $ - $ & $ - $ & $3.262$ \\ 
    & $28.672$ & $34.816$ & Band & $53.695$ & $5.000$ & $0.1193$ & $ - $ & $ - $ & $3.654$ \\ 
    & $34.816$ & $48.128$ & Band & $32.675$ & $5.000$ & $0.3055$ & $ - $ & $ - $ & $1.571$ \\ 
    \hline 
    \multirow{6}{*}{GRB120711A}
    & $48.128$ & $72.704$ & Band & $1801.327$ & $5.000$ & $0.0982$ & $ - $ & $1.277$ & $5.934$ \\ 
    & $72.704$ & $80.896$ & Band & $1351.595$ & $5.000$ & $0.1584$ & $ - $ & $ - $ & $2.751$ \\ 
    & $80.896$ & $90.112$ & Band & $350.544$ & $5.000$ & $0.0019$ & $44.793$ & $207.915$ & $ - $ \\ 
    & $90.112$ & $96.256$ & Band & $2596.171$ & $5.000$ & $0.0597$ & $ - $ & $1.733$ & $8.050$ \\ 
    & $96.256$ & $102.400$ & Band & $2342.635$ & $5.000$ & $0.0623$ & $ - $ & $1.659$ & $7.708$ \\ 
    & $102.400$ & $112.640$ & Band & $1652.340$ & $5.000$ & $0.1987$ & $ - $ & $ - $ & $2.416$ \\ 
    \hline 
    \multirow{1}{*}{GRB120716A}
    & $167.936$ & $190.464$ & Band & $115.188$ & $5.000$ & $0.3137$ & $ - $ & $ - $ & $2.107$ \\ 
    \hline 
    \multirow{1}{*}{GRB120909A}
    & $22.528$ & $59.392$ & CPL & $4287.832$ & $5.000$ & $0.3488$ & $ - $ & $ - $ & $2.298$ \\ 
    \hline 
    \multirow{1}{*}{GRB121128A}
    & $-2.048$ & $15.360$ & Band & $42.274$ & $5.000$ & $0.4014$ & $ - $ & $ - $ & $1.213$ \\ 
    \hline
    \multirow{12}{*}{GRB130427A}
    & $-2.048$ & $2.048$ & Band+BB & $363.389$ & $24.066$ & $0.0945$ & $2.037$ & $9.460$ & $ - $ \\ 
    & $2.048$ & $4.096$ & Band & $97.664$ & $5.000$ & $0.0527$ & $ - $ & $2.572$ & $11.948$ \\ 
    & $4.096$ & $5.120$ & Band+BB & $459.147$ & $33.025$ & $0.0771$ & $1.897$ & $8.814$ & $ - $ \\ 
    & $5.120$ & $6.144$ & Band+BB & $695.987$ & $39.277$ & $0.1016$ & $1.447$ & $6.722$ & $ - $ \\ 
    & $6.144$ & $7.168$ & Band+BB & $880.279$ & $44.453$ & $0.1132$ & $1.401$ & $6.510$ & $ - $ \\ 
    & $7.168$ & $8.192$ & Band+BB & $909.531$ & $48.543$ & $0.1168$ & $1.497$ & $6.955$ & $ - $ \\ 
    & $8.192$ & $9.216$ & Band+BB & $859.654$ & $41.976$ & $0.1053$ & $1.421$ & $6.598$ & $ - $ \\ 
    & $9.216$ & $10.240$ & Band+BB & $849.200$ & $40.612$ & $0.0889$ & $1.701$ & $7.902$ & $ - $ \\ 
    & $10.240$ & $11.264$ & Band+BB & $969.636$ & $33.876$ & $0.1075$ & $1.108$ & $5.149$ & $23.897$ \\ 
    & $11.264$ & $13.312$ & Band+BB & $525.769$ & $17.370$ & $0.0911$ & $ - $ & $4.191$ & $19.469$ \\ 
    & $13.312$ & $16.384$ & Band+BB & $362.916$ & $29.569$ & $0.0611$ & $5.270$ & $24.459$ & $ - $ \\ 
    & $16.384$ & $19.456$ & Band+BB & $257.621$ & $58.742$ & $0.1308$ & $4.602$ & $ - $ & $ - $ \\ 
    \hline 
    \multirow{5}{*}{GRB130518A}
    & $8.192$ & $22.528$ & Band & $358.915$ & $5.000$ & $0.0634$ & $ - $ & $3.005$ & $13.957$ \\ 
    & $22.528$ & $25.600$ & Band & $882.595$ & $5.000$ & $0.0722$ & $ - $ & $1.339$ & $6.220$ \\ 
    & $25.600$ & $28.672$ & CPL+BB & $426.850$ & $36.026$ & $0.0794$ & $ 3.231 $ & $15.007$ & $ - $ \\ 
    & $28.672$ & $33.792$ & CPL+BB & $172.195$ & $31.294$ & $0.0775$ & $ 3.815 $ & $17.721$ & $ - $ \\ 
    & $33.792$ & $47.104$ & Band+BB & $298.825$ & $22.358$ & $0.0959$ & $3.432$ & $15.941$ & $ - $ \\ 
    \hline 
    \multirow{2}{*}{GRB130610A}
    & $-52.224$ & $-19.456$ & Band+BB & $255.818$ & $30.196$ & $0.2863$ & $2.106$ & $ - $ & $ - $ \\ 
    & $-19.456$ & $4.096$ & CPL+BB & $232.275$ & $30.121$ & $0.3713$ & $1.303$ & $6.051$ & $ - $ \\ 
    \hline 
    \multirow{1}{*}{GRB130702A}
    & $-2.048$ & $13.312$ & CPL & $4219.994$ & $5.000$ & $0.2732$ & $ - $ & $1.122$ & $5.214$ \\ 
    \hline 
    \multirow{1}{*}{GRB131011A}
    & $-2.048$ & $18.432$ & CPL+BB & $277.374$ & $33.383$ & $0.1411$ & $3.533$ & $16.413$ & $ - $ \\ 
    \hline 
    \multirow{2}{*}{GRB131105A}
    & $16.384$ & $40.960$ & CPL & $3380.661$ & $5.000$ & $0.1977$ & $ - $ & $ - $ & $3.934$ \\ 
    & $79.872$ & $110.592$ & CPL & $355.126$ & $5.000$ & $0.3202$ & $ - $ & $ - $ & $2.203$ \\ 
    \hline 
    \multirow{2}{*}{GRB131108A}
    & $-7.168$ & $7.168$ & Band & $934.715$ & $5.000$ & $0.1726$ & $ - $ & $ - $ & $2.851$ \\ 
    & $7.168$ & $20.480$ & Band & $122.518$ & $10.000$ & $0.4831$ & $ - $ & $ - $ & $1.898$ \\ 
    \hline 
    \multirow{8}{*}{GRB131231A}
    & $-2.048$ & $14.336$ & Band & $904.124$ & $10.000$ & $0.4109$ & $ - $ & $ - $ & $3.327$ \\ 
    & $14.336$ & $19.456$ & Band & $873.872$ & $10.000$ & $0.4104$ & $ - $ & $ - $ & $1.863$ \\ 
    & $19.456$ & $22.528$ & Band & $357.972$ & $10.000$ & $0.1123$ & $ - $ & $1.960$ & $9.103$ \\ 
    & $22.528$ & $24.576$ & Band & $233.810$ & $10.000$ & $0.0601$ & $ - $ & $4.032$ & $18.730$ \\ 
    & $24.576$ & $27.648$ & Band & $131.167$ & $5.000$ & $0.0446$ & $ - $ & $2.648$ & $12.298$ \\ 
    & $27.648$ & $30.720$ & Band+BB & $46.439$ & $10.519$ & $0.0730$ & $ 1.028 $ & $ 4.777 $ & $ 22.172 $ \\ 
    & $30.720$ & $33.792$ & Band+BB & $313.132$ & $26.616$ & $0.2115$ & $ - $ & $4.112$ & $ - $ \\ 
    & $33.792$ & $38.912$ & Band & $37.503$ & $10.000$ & $0.2922$ & $ - $ & $ - $ & $3.500$ \\ 
    \hline
    \multirow{1}{*}{GRB140108A}
    & $74.752$ & $89.088$ & Band & $1019.830$ & $5.000$ & $0.1748$ & $ - $ & $ - $ & $4.283$ \\ 
    \hline
    \multirow{3}{*}{GRB140213A}
    & $-2.048$ & $4.096$ & Band & $170.056$ & $10.000$ & $0.2303$ & $ - $ & $1.026$ & $4.765$ \\ 
    & $4.096$ & $8.192$ & Band & $43.292$ & $20.000$ & $0.4561$ & $ - $ & $ - $ & $4.257$ \\ 
    & $8.192$ & $18.432$ & Band & $34.041$ & $5.000$ & $0.2557$ & $ - $ & $ - $ & $2.795$ \\ 
    \hline 
    \multirow{2}{*}{GRB140423A}
    & $-38.912$ & $-21.504$ & CPL+BB & $598.068$ & $39.178$ & $0.0318$ & $ 14.142 $ & $ 65.641 $ & $ - $ \\ 
    & $-21.504$ & $-1.024$ & Band & $2.309$ & $10.000$ & $0.0128$ & $ 7.676 $ & $ 35.656 $ & $ 165.502 $ \\ 
    \hline 
    \multirow{1}{*}{GRB140508A}
    & $-2.048$ & $12.288$ & Band+BB & $345.600$ & $31.571$ & $0.2286$ & $1.132$ & $5.257$ & $ - $ \\ 
    \hline 
    \multirow{2}{*}{GRB140512A}
    & $98.304$ & $120.832$ & CPL & $437.956$ & $5.000$ & $0.2413$ & $ - $ & $ - $ & $2.941$ \\ 
    & $120.832$ & $133.120$ & CPL & $293.555$ & $5.000$ & $0.2356$ & $ - $ & $ - $ & $2.481$ \\ 
    \hline
    \multirow{1}{*}{GRB140801A}
    & $-2.048$ & $7.168$ & Band & $83.642$ & $10.000$ & $0.1393$ & $ - $ & $2.251$ & $10.457$ \\ 
    \hline 
    \multirow{3}{*}{GRB141028A}
    & $-7.168$ & $13.312$ & Band & $1040.886$ & $5.000$ & $0.1681$ & $ - $ & $ - $ & $3.542$ \\ 
    & $13.312$ & $19.456$ & Band & $319.573$ & $5.000$ & $0.1203$ & $ - $ & $ - $ & $4.492$ \\ 
    & $19.456$ & $31.744$ & Band & $112.096$ & $5.000$ & $0.2936$ & $ - $ & $ - $ & $2.026$ \\ 
    \hline 
    \multirow{1}{*}{GRB141220A}
    & $-2.048$ & $7.168$ & Band & $156.070$ & $10.000$ & $0.2873$ & $ - $ & $ - $ & $4.604$ \\ 
    \hline  
    \multirow{3}{*}{GRB150314A}
    & $-2.048$ & $3.072$ & Band+BB & $305.738$ & $31.625$ & $0.0836$ & $2.567$ & $11.925$ & $ - $ \\ 
    & $3.072$ & $7.168$ & Band & $162.641$ & $20.000$ & $0.1280$ & $ - $ & $3.922$ & $18.205$ \\ 
    & $7.168$ & $14.336$ & Band & $511.829$ & $5.000$ & $0.1753$ & $ - $ & $ - $ & $2.484$ \\ 
    \hline 
    \multirow{4}{*}{GRB150403A}
    & $-2.048$ & $8.192$ & Band & $275.877$ & $30.911$ & $0.0526$ & $7.640$ & $35.462$ & $ - $ \\ 
    & $8.192$ & $12.288$ & Band & $672.910$ & $5.000$ & $0.0714$ & $ - $ & $1.461$ & $6.788$ \\ 
    & $12.288$ & $17.408$ & Band & $248.756$ & $20.000$ & $0.3460$ & $ - $ & $1.205$ & $5.600$ \\ 
    & $17.408$ & $29.696$ & Band & $188.314$ & $10.000$ & $0.3647$ & $ - $ & $ - $ & $3.644$ \\ 
    \hline 
    \multirow{4}{*}{GRB150821A}
    & $19.456$ & $26.624$ & Band & $117.491$ & $5.000$ & $0.2246$ & $ - $ & $ - $ & $2.339$ \\ 
    & $26.624$ & $35.840$ & Band & $277.098$ & $5.000$ & $0.2255$ & $ - $ & $ - $ & $2.566$ \\ 
    & $35.840$ & $44.032$ & Band & $482.127$ & $60.000$ & $0.2473$ & $2.726$ & $ - $ & $ - $ \\ 
    & $53.248$ & $66.560$ & Band & $105.527$ & $5.000$ & $0.4988$ & $ - $ & $ - $ & $1.053$ \\ 
    \hline 
    \multirow{2}{*}{GRB151027A}
    & $-2.048$ & $9.216$ & Band & $278.333$ & $43.259$ & $0.1597$ & $4.859$ & $ - $ & $ - $ \\ 
    & $45.056$ & $109.568$ & CPL & $2945.402$ & $5.000$ & $0.2663$ & $ - $ & $ - $ & $3.822$ \\ 
    \hline 
    \multirow{7}{*}{GRB160509A}
    & $-2.048$ & $10.240$ & CPL+BB & $321.717$ & $32.865$ & $0.1208$ & $ 3.236 $ & $15.032$ & $ - $ \\ 
    & $10.240$ & $12.288$ & Band & $343.113$ & $20.000$ & $0.4122$ & $ - $ & $ - $ & $2.913$ \\ 
    & $12.288$ & $14.336$ & Band & $396.505$ & $10.000$ & $0.0408$ & $1.253$ & $5.814$ & $27.008$ \\ 
    & $14.336$ & $16.384$ & Band & $209.444$ & $20.000$ & $0.1444$ & $ - $ & $2.637$ & $12.248$ \\ 
    & $16.384$ & $18.432$ & Band & $285.852$ & $20.000$ & $0.1329$ & $ - $ & $2.864$ & $13.303$ \\ 
    & $18.432$ & $22.528$ & Band & $200.546$ & $20.000$ & $0.3022$ & $ - $ & $1.260$ & $5.852$ \\ 
    & $22.528$ & $30.720$ & Band & $200.724$ & $5.000$ & $0.2293$ & $ - $ & $ - $ & $2.338$ \\ 
    \hline 
    \multirow{14}{*}{GRB160625B}
    & $178.176$ & $188.416$ & Band & $3123.807$ & $60.000$ & $0.0837$ & $8.116$ & $ - $ & $ - $ \\ 
    & $188.416$ & $189.440$ & Band+BB & $438.593$ & $47.696$ & $0.0621$ & $3.351$ & $15.564$ & $ - $ \\ 
    & $189.440$ & $191.488$ & Band+BB & $372.393$ & $33.625$ & $0.0578$ & $2.783$ & $12.927$ & $ - $ \\ 
    & $191.488$ & $193.536$ & Band & $471.893$ & $5.000$ & $0.0049$ & $7.446$ & $34.586$ & $160.532$ \\ 
    & $193.536$ & $194.560$ & Band+BB & $368.194$ & $24.159$ & $0.0241$ & $6.018$ & $27.956$ & $ - $ \\ 
    & $194.560$ & $196.608$ & Band+BB & $317.493$ & $29.021$ & $0.0329$ & $5.035$ & $23.387$ & $ - $ \\ 
    & $196.608$ & $198.656$ & Band+BB & $280.552$ & $17.902$ & $0.0250$ & $4.235$ & $19.670$ & $91.301$ \\ 
    & $198.656$ & $199.680$ & Band+BB & $373.748$ & $18.802$ & $0.0231$ & $4.862$ & $22.582$ & $104.817$ \\ 
    & $199.680$ & $200.704$ & Band & $969.729$ & $5.000$ & $0.0009$ & $51.755$ & $240.226$ & $1115.029$ \\ 
    & $200.704$ & $202.752$ & Band+BB & $520.856$ & $27.101$ & $0.0178$ & $10.277$ & $47.701$ & $ - $ \\ 
    & $202.752$ & $204.800$ & Band & $345.542$ & $40.000$ & $0.0184$ & $20.683$ & $96.000$ & $ - $ \\ 
    & $204.800$ & $206.848$ & Band & $246.471$ & $5.000$ & $0.0094$ & $3.481$ & $16.167$ & $75.042$ \\ 
    & $206.848$ & $209.920$ & Band & $116.407$ & $10.000$ & $0.1109$ & $ - $ & $1.914$ & $8.892$ \\ 
    & $214.016$ & $222.208$ & Band & $822.360$ & $5.000$ & $0.1938$ & $ - $ & $ - $ & $2.728$ \\ 
    \hline 
    \multirow{1}{*}{GRB160629A}
    & $15.360$ & $28.672$ & Band & $856.776$ & $50.000$ & $0.3694$ & $1.647$ & $ - $ & $ - $ \\ 
    \hline 
    \multirow{1}{*}{GRB160804A}
    & $-24.576$ & $0.000$ & Band & $42.214$ & $5.000$ & $0.4480$ & $ - $ & $ - $ & $1.577$ \\ 
    \hline
    \multirow{1}{*}{GRB161117A}
    & $36.864$ & $47.104$ & Band & $93.733$ & $10.000$ & $0.5017$ & $ - $ & $ - $ & $2.160$ \\ 
    \hline 
    \multirow{13}{*}{GRB170214A}
    & $-2.048$ & $12.288$ & Band+BB & $2389.995$ & $32.918$ & $0.0457$ & $ 13.218 $ & $ - $ & $ - $ \\ 
    & $12.288$ & $20.480$ & Band & $272.135$ & $5.000$ & $0.1597$ & $ - $ & $ - $ & $3.105$ \\ 
    & $20.480$ & $28.672$ & Band+BB & $320.507$ & $33.556$ & $0.1446$ & $2.390$ & $11.102$ & $ - $ \\ 
    & $28.672$ & $36.864$ & Band+BB & $2413.741$ & $32.896$ & $0.0332$ & $ 15.696 $ & $ 72.855 $ & $ - $ \\ 
    & $36.864$ & $44.032$ & Band+BB & $340.073$ & $46.585$ & $0.1635$ & $2.604$ & $12.095$ & $ - $ \\ 
    & $44.032$ & $50.176$ & Band+BB & $1449.851$ & $41.194$ & $0.0998$ & $ 4.009 $ & $ 18.621 $ & $ - $ \\ 
    & $50.176$ & $57.344$ & Band+BB & $347.149$ & $37.892$ & $0.1337$ & $2.686$ & $12.477$ & $ - $ \\ 
    & $57.344$ & $63.488$ & Band & $1087.015$ & $5.000$ & $0.1337$ & $ - $ & $ - $ & $3.367$ \\ 
    & $63.488$ & $69.632$ & Band+BB & $2152.000$ & $31.548$ & $0.0262$ & $ 17.838 $ & $ 82.794 $ & $ - $ \\ 
    & $69.632$ & $76.800$ & Band & $1900.238$ & $5.000$ & $0.0227$ & $1.996$ & $9.271$ & $43.032$ \\ 
    & $107.520$ & $128.000$ & Band & $3580.956$ & $5.000$ & $0.3316$ & $ - $ & $ - $ & $1.998$ \\ 
    & $128.000$ & $137.216$ & CPL+BB & $230.044$ & $32.561$ & $0.0987$ & $3.959$ & $18.387$ & $ - $ \\ 
    & $137.216$ & $148.480$ & Band & $3576.749$ & $5.000$ & $0.2901$ & $ - $ & $ - $ & $1.883$ \\ 
    \hline 
    \multirow{3}{*}{GRB170405A}
    & $-2.048$ & $22.528$ & Band & $1142.968$ & $5.000$ & $0.2301$ & $ - $ & $ - $ & $2.531$ \\ 
    & $22.528$ & $34.816$ & Band & $770.655$ & $5.000$ & $0.1888$ & $ - $ & $ - $ & $2.801$ \\ 
    & $34.816$ & $53.248$ & Band & $66.028$ & $5.000$ & $0.1734$ & $ - $ & $ - $ & $3.704$ \\ 
    \hline 
    GRB171010A & $-2.048$ & $11.264$ & Band & $113.691$ & $5.000$ & $0.2126$ & $ - $ & $ - $ & $3.642$ \\ 
    & $11.264$ & $15.360$ & Band & $647.195$ & $5.000$ & $0.0537$ & $ - $ & $3.105$ & $14.422$ \\ 
    & $15.360$ & $18.432$ & Band & $142.542$ & $5.000$ & $0.0650$ & $ - $ & $2.329$ & $10.818$ \\ 
    & $18.432$ & $20.480$ & Band & $110.204$ & $5.000$ & $0.0212$ & $1.866$ & $8.669$ & $40.237$ \\ 
    & $20.480$ & $22.528$ & Band & $131.231$ & $5.000$ & $0.0175$ & $2.265$ & $10.522$ & $48.837$ \\ 
    & $22.528$ & $25.600$ & Band & $107.062$ & $5.000$ & $0.0226$ & $1.755$ & $8.151$ & $37.832$ \\ 
    & $25.600$ & $28.672$ & Band & $102.489$ & $5.000$ & $0.0203$ & $2.152$ & $9.997$ & $46.403$ \\ 
    & $28.672$ & $30.720$ & Band & $110.762$ & $5.000$ & $0.0282$ & $1.275$ & $5.916$ & $27.480$ \\ 
    & $30.720$ & $32.768$ & Band+BB & $431.974$ & $29.396$ & $0.0524$ & $5.025$ & $23.340$ & $ - $ \\ 
    & $32.768$ & $34.816$ & Band+BB & $293.178$ & $11.267$ & $0.0411$ & $1.831$ & $8.507$ & $39.486$ \\ 
    & $34.816$ & $36.864$ & Band & $147.281$ & $5.000$ & $0.0088$ & $5.434$ & $25.242$ & $117.164$ \\ 
    & $36.864$ & $38.912$ & Band & $62.595$ & $5.000$ & $0.0240$ & $1.652$ & $7.672$ & $35.610$ \\ 
    & $38.912$ & $41.984$ & Band+BB & $242.835$ & $29.220$ & $0.0934$ & $3.286$ & $15.263$ & $ - $ \\ 
    & $41.984$ & $45.056$ & Band & $62.341$ & $5.000$ & $0.0256$ & $1.703$ & $7.910$ & $36.714$ \\ 
    & $45.056$ & $48.128$ & Band & $68.396$ & $5.000$ & $0.0215$ & $2.032$ & $9.440$ & $43.816$ \\ 
    & $48.128$ & $51.200$ & Band & $41.361$ & $10.000$ & $0.0659$ & $1.496$ & $6.947$ & $32.245$ \\ 
    & $51.200$ & $54.272$ & Band & $73.462$ & $5.000$ & $0.0327$ & $1.211$ & $5.620$ & $26.106$ \\ 
    & $54.272$ & $56.320$ & Band & $165.010$ & $5.000$ & $0.0207$ & $1.912$ & $8.880$ & $41.217$ \\ 
    & $56.320$ & $58.368$ & Band+BB & $428.130$ & $10.484$ & $0.0264$ & $3.535$ & $16.418$ & $76.205$ \\ 
    & $58.368$ & $60.416$ & Band+BB & $238.891$ & $8.453$ & $0.0473$ & $1.155$ & $5.362$ & $24.905$ \\ 
    & $60.416$ & $63.488$ & Band & $90.191$ & $5.000$ & $0.0225$ & $1.760$ & $8.174$ & $37.942$ \\ 
    & $63.488$ & $65.536$ & Band & $86.954$ & $5.000$ & $0.0247$ & $1.601$ & $7.432$ & $34.520$ \\ 
    & $65.536$ & $67.584$ & Band & $79.856$ & $5.000$ & $0.0144$ & $3.027$ & $14.060$ & $65.261$ \\ 
    & $67.584$ & $69.632$ & Band & $85.106$ & $5.000$ & $0.0098$ & $4.901$ & $22.766$ & $105.670$ \\ 
    & $69.632$ & $71.680$ & Band & $103.378$ & $5.000$ & $0.0121$ & $3.974$ & $18.458$ & $85.677$ \\ 
    & $71.680$ & $74.752$ & Band & $65.037$ & $20.000$ & $0.0518$ & $4.089$ & $18.994$ & $ - $ \\ 
    & $74.752$ & $77.824$ & Band+BB & $234.017$ & $29.523$ & $0.1561$ & $1.729$ & $8.032$ & $ - $ \\ 
    & $77.824$ & $81.920$ & Band+BB & $195.867$ & $25.349$ & $0.1258$ & $2.274$ & $10.563$ & $ - $ \\ 
    & $81.920$ & $84.992$ & Band+BB & $246.436$ & $35.160$ & $0.1327$ & $2.830$ & $13.144$ & $ - $ \\ 
    & $84.992$ & $89.088$ & Band+BB & $367.023$ & $29.634$ & $0.1105$ & $3.545$ & $16.466$ & $ - $ \\ 
    & $89.088$ & $95.232$ & Band & $43.240$ & $5.000$ & $0.1962$ & $ - $ & $ - $ & $3.583$ \\ 
    & $110.592$ & $118.784$ & Band & $34.079$ & $5.000$ & $0.2292$ & $ - $ & $ - $ & $3.380$ \\ 
    & $128.000$ & $136.192$ & Band & $29.093$ & $5.000$ & $0.1796$ & $ - $ & $1.023$ & $4.750$ \\ 
    & $136.192$ & $144.384$ & Band & $30.314$ & $5.000$ & $0.2910$ & $ - $ & $ - $ & $2.415$ \\ 
    \hline 
    \multirow{1}{*}{GRB180620B}
    & $-7.168$ & $14.336$ & Band & $113.441$ & $5.000$ & $0.4164$ & $ - $ & $ - $ & $1.294$ \\ 
    \hline
    \multirow{1}{*}{GRB180703A}
    & $-7.168$ & $10.240$ & Band & $305.096$ & $26.586$ & $0.0643$ & $6.569$ & $30.491$ & $ - $ \\ 
    \hline
    \multirow{10}{*}{GRB180720B}
    & $-7.168$ & $6.144$ & Band & $1631.934$ & $5.000$ & $0.0736$ & $ - $ & $1.937$ & $8.995$ \\ 
    & $6.144$ & $9.216$ & Band & $2376.161$ & $5.000$ & $0.0272$ & $1.024$ & $4.752$ & $22.072$ \\ 
    & $9.216$ & $11.264$ & Band & $2401.166$ & $5.000$ & $0.0238$ & $1.171$ & $5.438$ & $25.258$ \\ 
    & $11.264$ & $14.336$ & Band+BB & $271.121$ & $29.747$ & $0.0449$ & $5.487$ & $25.487$ & $ - $ \\ 
    & $14.336$ & $16.384$ & Band+BB & $277.642$ & $26.802$ & $0.0361$ & $5.455$ & $25.336$ & $ - $ \\ 
    & $16.384$ & $18.432$ & Band+BB & $474.485$ & $28.498$ & $0.0276$ & $7.891$ & $36.627$ & $ - $ \\ 
    & $18.432$ & $22.528$ & Band & $133.121$ & $5.000$ & $0.0372$ & $ - $ & $3.831$ & $17.793$ \\ 
    & $26.624$ & $29.696$ & Band & $125.858$ & $60.000$ & $0.0784$ & $8.937$ & $ - $ & $ - $ \\ 
    & $29.696$ & $35.840$ & Band & $817.507$ & $5.000$ & $0.0793$ & $ - $ & $1.632$ & $7.581$ \\ 
    & $47.104$ & $52.224$ & Band & $650.776$ & $10.000$ & $0.2550$ & $ - $ & $ - $ & $3.989$ \\ 
    \hline 
    \multirow{3}{*}{GRB180728A}
    & $-7.168$ & $11.264$ & Band+BB & $32.534$ & $4.725$ & $0.0782$ & $ 1.183 $ & $ 5.495 $ & $ 25.504 $ \\ 
    & $11.264$ & $13.312$ & Band & $62.686$ & $5.000$ & $0.0225$ & $2.729$ & $12.676$ & $58.839$ \\ 
    & $13.312$ & $17.408$ & Band & $39.554$ & $5.000$ & $0.0805$ & $ - $ & $3.906$ & $18.145$ \\ 
    \hline 
    \multirow{1}{*}{GRB181020A}
    & $-2.048$ & $6.144$ & Band & $1330.896$ & $5.000$ & $0.1082$ & $ - $ & $1.229$ & $5.711$ \\ 
    & $6.144$ & $12.288$ & Band & $635.015$ & $5.000$ & $0.0805$ & $ - $ & $1.651$ & $7.668$ \\ 
    \hline 
    \multirow{8}{*}{GRB190114C}
    & $-2.048$ & $1.024$ & CPL & $664.801$ & $10.000$ & $0.0582$ & $1.038$ & $4.819$ & $22.386$ \\ 
    & $1.024$ & $2.048$ & CPL+BB & $404.340$ & $12.120$ & $0.0095$ & $8.628$ & $40.078$ & $186.027$ \\ 
    & $2.048$ & $3.072$ & CPL & $1095.919$ & $60.000$ & $0.0263$ & $19.111$ & $88.706$ & $ - $ \\ 
    & $3.072$ & $4.096$ & Band+BB & $676.468$ & $12.187$ & $0.0117$ & $5.656$ & $26.273$ & $121.950$ \\ 
    & $4.096$ & $6.144$ & CPL & $639.452$ & $20.000$ & $0.0090$ & $19.249$ & $89.346$ & $ - $ \\ 
    & $6.144$ & $11.264$ & CPL & $4290.623$ & $60.000$ & $0.2884$ & $2.112$ & $ - $ & $ - $ \\ 
    & $11.264$ & $17.408$ & Band+BB & $241.065$ & $13.750$ & $0.0847$ & $1.718$ & $7.980$ & $ - $ \\ 
    & $17.408$ & $22.528$ & CPL & $32.926$ & $5.000$ & $0.1774$ & $ - $ & $ - $ & $3.579$ \\ 
    \hline 
    \multirow{2}{*}{GRB190324A}
    & $-2.048$ & $23.552$ & Band & $140.940$ & $10.000$ & $0.5164$ & $ - $ & $ - $ & $2.132$ \\ 
    & $23.552$ & $33.792$ & Band & $72.268$ & $5.000$ & $0.2236$ & $ - $ & $ - $ & $2.911$ \\ 
    \hline 
    \multirow{2}{*}{GRB190829A}
    & $37.888$ & $53.248$ & Band & $8.944$ & $5.000$ & $0.2099$ & $ - $ & $2.303$ & $10.696$ \\ 
    & $53.248$ & $59.392$ & Band & $12.306$ & $5.000$ & $0.1160$ & $ - $ & $4.594$ & $21.322$ \\ 
    \hline
    \multirow{2}{*}{GRB200524A}
    & $-2.048$ & $8.192$ & Band+BB & $4007.341$ & $22.658$ & $0.0917$ & $ 2.642 $ & $ 12.272 $ & $ - $ \\ 
    & $8.192$ & $21.504$ & Band+BB & $375.375$ & $17.078$ & $0.0830$ & $2.664$ & $12.374$ & $ - $ \\ 
    \hline 
    \multirow{3}{*}{GRB200613A}
    & $-2.048$ & $7.168$ & Band & $108.873$ & $10.000$ & $0.2856$ & $ - $ & $ - $ & $4.228$ \\ 
    & $7.168$ & $12.288$ & Band & $71.854$ & $10.000$ & $0.3125$ & $ - $ & $ - $ & $2.890$ \\ 
    & $12.288$ & $16.384$ & Band & $101.548$ & $20.000$ & $0.3052$ & $ - $ & $1.507$ & $6.998$ \\ 
    \hline
    \multirow{1}{*}{GRB200826A}
    & $0.000$ & $0.500$ & CPL+BB & $151$ & $23.666$ & $0.0015$ & $759.355$ & $ - $ & $ - $ \\ 
    \hline
    \multirow{4}{*}{GRB200829A}
    & $-2.048$ & $18.432$ & Band & $302.776$ & $5.000$ & $0.1814$ & $ - $ & $ - $ & $2.665$ \\ 
    & $18.432$ & $20.480$ & Band+BB & $359.650$ & $32.662$ & $0.0745$ & $1.760$ & $8.174$ & $ - $ \\ 
    & $20.480$ & $23.552$ & Band+BB & $378.778$ & $27.293$ & $0.0863$ & $1.452$ & $6.746$ & $31.313$ \\ 
    & $23.552$ & $38.912$ & Band+BB & $120.767$ & $29.659$ & $0.1125$ & $ 3.761 $ & $ 17.471 $ & $ - $ \\ 
    \hline 
    \multirow{2}{*}{GRB201020B}
    & $-2.048$ & $10.240$ & CPL+BB & $591.720$ & $22.627$ & $0.1547$ & $1.329$ & $6.174$ & $ - $ \\ 
    & $10.240$ & $17.408$ & Band & $57.741$ & $10.000$ & $0.2761$ & $ - $ & $ - $ & $3.496$ \\ 
    \hline 
    \multirow{1}{*}{GRB201104C}
    & $-2.048$ & $27.648$ & Band & $2.067$ & $5.000$ & $0.3917$ & $ - $ & $ - $ & $2.234$ \\ 
    \hline 
    \multirow{8}{*}{GRB201216C}
    & $-7.168$ & $8.192$ & Band & $1226.549$ & $5.000$ & $0.0988$ & $ - $ & $1.428$ & $6.633$ \\ 
    & $8.192$ & $12.288$ & Band & $170.307$ & $10.000$ & $0.1504$ & $ - $ & $1.440$ & $6.691$ \\ 
    & $12.288$ & $17.408$ & Band & $110.066$ & $5.000$ & $0.0852$ & $ - $ & $1.503$ & $6.983$ \\ 
    & $17.408$ & $21.504$ & Band & $119.054$ & $10.000$ & $0.1512$ & $ - $ & $1.433$ & $6.655$ \\ 
    & $21.504$ & $24.576$ & Band & $127.093$ & $10.000$ & $0.1170$ & $ - $ & $1.852$ & $8.603$ \\ 
    & $24.576$ & $26.624$ & Band & $437.583$ & $10.000$ & $0.0555$ & $ - $ & $4.304$ & $19.994$ \\ 
    & $26.624$ & $29.696$ & Band & $353.328$ & $5.000$ & $0.0209$ & $1.455$ & $6.755$ & $31.375$ \\ 
    & $29.696$ & $34.816$ & Band & $79.439$ & $10.000$ & $0.2306$ & $ - $ & $1.035$ & $4.808$ \\ 
    \hline
    \multirow{5}{*}{GRB210204A}
    & $178.176$ & $192.512$ & Band & $157.621$ & $5.000$ & $0.2408$ & $ - $ & $ - $ & $2.451$ \\ 
    & $192.512$ & $200.704$ & Band & $59.173$ & $5.000$ & $0.2732$ & $ - $ & $ - $ & $1.617$ \\ 
    & $200.704$ & $204.800$ & Band & $57.457$ & $10.000$ & $0.3064$ & $ - $ & $ - $ & $2.688$ \\ 
    & $210.944$ & $217.088$ & Band & $45.935$ & $5.000$ & $0.2300$ & $ - $ & $ - $ & $1.920$ \\ 
    & $217.088$ & $223.232$ & Band & $47.216$ & $5.000$ & $0.2626$ & $ - $ & $ - $ & $1.682$ \\ 
    \hline 
    \multirow{5}{*}{GRB210610B}
     & $-2.048$ & $26.624$ & mBB & $190.347$ & $-0.134$ & $0.2101$ & $ - $ & $ - $ & $2.821$ \\ 
     & $26.624$ & $32.768$ & mBB & $951.700$ & $-0.276$ & $0.1121$ & $ - $ & $ - $ & $3.591$ \\ 
     & $32.768$ & $39.936$ & mBB & $344.822$ & $-0.151$ & $0.2875$ & $ - $ & $ - $ & $2.876$ \\ 
     & $39.936$ & $48.128$ & mBB & $112.503$ & $-1.882$ & $0.1677$ & $ - $ & $ - $ & $2.645$ \\ 
     & $48.128$ & $60.416$ & mBB & $68.578$ & $-1.547$ & $0.4964$ & $ - $ & $ - $ & $2.020$ \\ 
    \hline
    \multirow{1}{*}{GRB210704A}
    & $-7.168$ & $4.096$ & CPL & $659.428$ & $5.000$ & $0.1841$ & $ - $ & $ - $ & $2.624$ \\ 
    \hline 
    \multirow{4}{*}{GRB211018A}
    & $-2.048$ & $9.216$ & Band & $397.962$ & $10.000$ & $0.4916$ & $ - $ & $ - $ & $1.889$ \\ 
    & $9.216$ & $19.456$ & Band & $269.487$ & $5.000$ & $0.1661$ & $ - $ & $ - $ & $3.642$ \\ 
    & $34.816$ & $49.152$ & Band & $1093.520$ & $5.000$ & $0.1905$ & $ - $ & $ - $ & $3.177$ \\ 
    & $49.152$ & $60.416$ & Band & $1171.723$ & $5.000$ & $0.1814$ & $ - $ & $ - $ & $3.025$ \\ 
    & $74.752$ & $82.944$ & Band & $1511.188$ & $5.000$ & $0.1275$ & $ - $ & $ - $ & $4.307$ \\ 
    & $82.944$ & $92.160$ & Band & $689.554$ & $5.000$ & $0.1526$ & $ - $ & $ - $ & $3.600$ \\ 
    \hline 
    \multirow{3}{*}{GRB211023A}
    & $61.440$ & $66.560$ & Band & $55.644$ & $5.000$ & $0.2053$ & $ - $ & $ - $ & $2.382$ \\ 
    & $66.560$ & $71.680$ & Band & $55.427$ & $20.000$ & $0.1872$ & $ - $ & $4.026$ & $ - $ \\ 
    & $83.968$ & $89.088$ & Band & $26.582$ & $5.000$ & $0.1106$ & $ - $ & $1.546$ & $7.180$ \\ 
    \hline 
    \multirow{15}{*}{GRB211211A}
    & $-2.048$ & $3.072$ & Band+BB & $305.909$ & $32.429$ & $0.0521$ & $12.063$ & $ - $ & $ - $ \\ 
    & $3.072$ & $5.120$ & Band+BB & $481.812$ & $23.429$ & $0.0734$ & $3.336$ & $15.497$ & $ - $ \\ 
    & $5.120$ & $6.144$ & Band+BB & $404.987$ & $25.219$ & $0.0723$ & $3.579$ & $16.623$ & $ - $ \\ 
    & $6.144$ & $7.168$ & Band+BB & $862.219$ & $33.844$ & $0.1160$ & $2.215$ & $10.288$ & $ - $ \\ 
    & $7.168$ & $9.216$ & Band+BB & $939.338$ & $28.202$ & $0.0910$ & $2.706$ & $12.568$ & $ - $ \\ 
    & $9.216$ & $11.264$ & Band+BB & $243.001$ & $11.269$ & $0.0654$ & $2.106$ & $9.780$ & $ - $ \\ 
    & $11.264$ & $16.384$ & Band & $61.088$ & $5.000$ & $0.0793$ & $ - $ & $3.769$ & $17.508$ \\ 
    & $16.384$ & $18.432$ & Band & $91.632$ & $20.000$ & $0.1148$ & $2.244$ & $10.424$ & $ - $ \\ 
    & $18.432$ & $20.480$ & Band+BB & $123.170$ & $17.379$ & $0.0978$ & $ 2.151 $ & $ 9.991 $ & $ - $ \\ 
    & $20.480$ & $22.528$ & Band & $83.240$ & $5.000$ & $0.0232$ & $3.058$ & $14.203$ & $65.924$ \\ 
    & $22.528$ & $24.576$ & Band & $101.978$ & $5.000$ & $0.0191$ & $3.724$ & $17.300$ & $80.298$ \\ 
    & $24.576$ & $27.648$ & Band+BB & $370.581$ & $31.578$ & $0.0973$ & $6.003$ & $ - $ & $ - $ \\ 
    & $31.744$ & $35.840$ & Band & $45.842$ & $5.000$ & $0.1482$ & $ - $ & $1.662$ & $7.722$ \\ 
    & $35.840$ & $39.936$ & Band+BB & $309.212$ & $35.714$ & $0.0874$ & $10.366$ & $ - $ & $ - $ \\ 
    & $51.200$ & $59.392$ & Band & $30.446$ & $5.000$ & $0.2744$ & $ - $ & $1.090$ & $5.062$ \\ 
    \hline 
    \multirow{3}{*}{GRB220101A}
    & $93.184$ & $102.400$ & Band & $310.882$ & $10.000$ & $0.4839$ & $ - $ & $ - $ & $1.995$ \\ 
    & $102.400$ & $110.592$ & Band & $463.488$ & $5.000$ & $0.1403$ & $ - $ & $ - $ & $4.485$ \\ 
    & $110.592$ & $119.808$ & Band & $166.610$ & $5.000$ & $0.1579$ & $ - $ & $ - $ & $4.387$ \\ 
    \hline 
    \multirow{1}{*}{GRB220107A}
    & $-2.048$ & $10.240$ & Band & $256.505$ & $5.000$ & $0.2435$ & $ - $ & $ - $ & $2.189$ \\ 
    \hline 
    \multirow{4}{*}{GRB220527A}
    & $-2.048$ & $4.096$ & Band & $93.833$ & $60.000$ & $0.1453$ & $4.512$ & $ - $ & $ - $ \\ 
    & $4.096$ & $7.168$ & Band & $97.901$ & $5.000$ & $0.0311$ & $ - $ & $4.294$ & $19.946$ \\ 
    & $7.168$ & $11.264$ & Band+BB & $221.878$ & $34.365$ & $0.2607$ & $ - $ & $3.611$ & $ - $ \\ 
    & $11.264$ & $23.552$ & Band & $843.126$ & $5.000$ & $0.2946$ & $ - $ & $ - $ & $2.103$ \\ 
    \hline 
    \multirow{5}{*}{GRB220627A}
    & $34.816$ & $59.392$ & Band & $369.824$ & $5.000$ & $0.2217$ & $ - $ & $ - $ & $3.761$ \\ 
    & $82.944$ & $101.376$ & Band & $196.188$ & $5.000$ & $0.3484$ & $ - $ & $ - $ & $1.789$ \\ 
    & $124.928$ & $147.456$ & Band & $1797.749$ & $10.000$ & $0.5087$ & $ - $ & $ - $ & $2.284$ \\ 
    & $147.456$ & $161.792$ & Band & $261.076$ & $10.000$ & $0.4308$ & $ - $ & $ - $ & $2.697$ \\ 
    & $161.792$ & $178.176$ & Band & $5000.000$ & $30.000$ & $0.0005$ & $ - $ & $ - $ & $ - $ \\ 
    \hline 
\end{longtable}

\begin{deluxetable}{c|c}[htp]
    \label{table_eq1}
    \caption{Non-dissipation regime criteria}
    \tablehead{\colhead{Regime} & \colhead{Criteria} }
    \startdata
    Regime II & $14.8(1+z)^{1/4}(\frac{kT_{ob}}{50keV})^{1/4}(\frac{F_{BB}}{10^{-8}erg\ s^{-1}cm^{-2}})^{3/16}r_{0,9}^{1/8}f_{th,-1}^{1/2}f_{\gamma}^{1/2}d_{L,28}^{3/8}>1$ \\
    $\eta>(1+\sigma_{0})^{1/2}$ & $0.24(1+z)^{-3}(\frac{kT_{ob}}{50keV})^{-3}(\frac{F_{BB}}{10^{-8}erg\ s^{-1}cm^{-2}})^{3/4}r_{0,9}^{-3/2}d_{L,28}^{3/2}>1$ \\
    $r_{c}>r_{ph}>r_{ra}$ & $1.43\times 10^{-5}(1+z)^{-7}(\frac{kT_{ob}}{50keV})^{-7}(\frac{F_{BB}}{10^{-8}erg\ s^{-1}cm^{-2}})^{7/4}r_{0,9}^{-7/2}f_{th,-1}^{3}f_{\gamma}^{3}d_{L,28}^{7/2}<1$ \\
    \hline
    Regime III & $8.28(1+z)^{-3/2}(\frac{kT_{ob}}{30keV})^{-3/2}(\frac{F_{BB}}{10^{-7}erg\ s^{-1}cm^{-2}})^{5/8}r_{0,9}^{-1}f_{th,-1}^{5/4}f_{\gamma}^{5/4}d_{L,28}^{5/4}>1$ \\
    $\eta>(1+\sigma_{0})^{1/2}$ & $9.42\times 10^{-2}(1+z)^{-14/3}(\frac{kT_{ob}}{30keV})^{-14/3}(\frac{F_{BB}}{10^{-7}erg\ s^{-1}cm^{-2}})^{7/6}r_{0,9}^{-7/3}f_{th,-1}^{2}f_{\gamma}^{2}d_{L,28}^{7/3}>1$ \\
    $r_{ph}>r_{c}$ &  \\
    \hline
    Regime V & $41.4(1+z)^{1/2}(\frac{kT_{ob}}{10keV})^{1/2}(\frac{F_{BB}}{10^{-9}erg\ s^{-1}cm^{-2}})^{3/8}r_{0,9}^{-1/4}f_{th,-1}f_{\gamma}d_{L,28}^{3/4}<1$ \\
    $\eta<(1+\sigma_{0})^{1/2}$ & $5.28(1+z)^{-3}(\frac{kT_{ob}}{10keV})^{-3}(\frac{F_{BB}}{10^{-9}erg\ s^{-1}cm^{-2}})^{3/4}r_{0,9}^{-3/2}d_{L,28}^{3/2}>1$ \\
    $r_{c}>r_{ph}>r_{ra}$ & $1.16\times 10^{-5}(1+z)^{-8}(\frac{kT_{ob}}{10keV})^{-8}(\frac{F_{BB}}{10^{-9}erg\ s^{-1}cm^{-2}})r_{0,9}^{-3}f_{th,-1}f_{\gamma}d_{L,28}^{2}<1$ \\
    \hline
    Regime VI & $8.28(1+z)^{-3/2}(\frac{kT_{ob}}{30keV})^{-3/2}(\frac{F_{BB}}{10^{-7}erg\ s^{-1}cm^{-2}})^{5/8}r_{0,9}^{-1}f_{th,-1}^{5/4}f_{\gamma}^{5/4}d_{L,28}^{5/4}<1$ \\
    $\eta<(1+\sigma_{0})^{1/2}$ & $5.63\times 10^{-3}(1+z)^{-8/3}(\frac{kT_{ob}}{30keV})^{-8/3}(\frac{F_{BB}}{10^{-7}erg\ s^{-1}cm^{-2}})^{1/3}r_{0,9}^{-1}f_{th,-1}^{1/3}f_{\gamma}^{1/3}d_{L,28}^{2/3}>1$ \\
    $r_{ph}>r_{c}$ &  \\
    \enddata
\end{deluxetable}
\vspace{0.1ex}

\end{ThreePartTable}

\end{document}